# Non-Hermiticities even in quantum systems that are closed

K. Moulopoulos

Department of Physics, University of Cyprus, 75 Kallipoleos Str., Nicosia, 1678, Cyprus

**Abstract:** Rarely noted paradoxes in applications of fundamental quantum relations are pointed out, with their resolution leading to emergent non-Hermitian behaviors due to boundary terms – even for closed systems and with real potentials. The role played by these non-Hermiticities on the consistency of quantum mechanical uncertainty relations is discussed, especially in multiply-connected spaces (more generally, for any system that satisfies the Bloch theorem of Solid State Physics). These subtleties – reflections of topological quantum anomalies – follow their own patterns (for any dimensionality, for both Schrödinger and Dirac/Weyl Hamiltonians and for either continuous or lattice (tight-binding) models)**:** they can always be written as global fluxes of certain generalized current densities $\mathbf{J}_g$. In continuous nonrelativistic models, these have the forms that had earlier been used by Chemists to describe atomic fragments of polyatomic molecules, while for Dirac/Weyl or other lattice models they have more interesting relativistic forms only recently worked out in graphene models. In spite of the deep mathematical origin as quantum anomalies examples are provided here, where such non-Hermiticities have a direct physical significance (for both conventional and topological materials). In all stationary state examples considered, these non-Hermitian boundary terms turn out to be quantized, this quantization being either of conventional or of a topological (Quantum Hall Effect (QHE)-type) origin. The latter claim is substantiated through direct application to a simple QHE arrangement (2D Landau system in an external in-plane electric field), where some particular $\mathbf{J}_g$ seems to be related to the well-known dissipationless edge currents. More generally, the non-Hermitian terms play a subtle role on Berry curvatures in solids and seem to be crucial for the consistent application of the so called Modern Theories of Polarization and Orbital Magnetization. It is emphasized that the above systems can be *closed* (in multiply-connected space, so that the boundaries disappear, but the non-Hermiticity remains), a case in non-Hermitian physics that has never been discussed in the literature**;** it is also stressed that a mapping between the above non-Hermiticity (for continuous systems) and the many recent available results in tight-binding solid state models (leading to the so-called exceptional points) will lead to enhanced understanding of quantum behavior at the most fundamental level.

This note is a perspective on concepts of non-Hermiticity that are not usually discussed. It gives an overview of recent work - but also an outlook on future possibilities - related to earlier paradoxes that, for real potentials, originate from hidden non-Hermiticity of the Hamiltonian (due to the kinetic energy operator) of any quantum system**;** and, to make it more dramatic, we confine ourselves to simple *closed* systems, where total probability is conserved – unlike the typical case of non-Hermitian models discussed in the literature (involving open systems). The catch is spatial multiple-connectedness (when the edges of the system disappear due to gluing and the environment is identical to the bulk of the system itself), and this is emphasized here for the first time. Apart from this novelty, we remind the reader that the above mentioned paradoxes had been noted in applications of the Ehrenfest theorem and Hellmann-Feynman theorems, with some related but separate discussions on the quantum mechanical uncertainty relations. These few works were totally disconnected to each other and the whole issue has been largely ignored, until recently – when a new analysis of the matter seems to lead to interesting possibilities.

The paradoxes are resolved if the proper boundary terms resulting from certain integration by parts (almost always discarded in the literature) are retained and are studied seriously. [One should point out, however, that there are a few very recent works that have started seriously discussing those boundary terms, see i.e. ref. [24] for a pedagogical analysis.] These extra boundary terms (once again reflections of non-Hermiticities, but at a deeper level of topological anomalies) seem to follow their own behavioral patterns (for systems of any dimensionality, for both Schrödinger and Dirac/Weyl Hamiltonians and for both continuous and lattice (tight-binding) models)**:** they can always be written as global fluxes of certain generalized current densities $\mathbf{J}_g^{\mathbf{\Omega}}$ across the system boundaries, and these $\mathbf{J}_g^{\mathbf{\Omega}}$ are defined through the use of any input vector operator $\mathbf{\Omega}$ (the one that has been used as input i.e. in the corresponding Ehrenfest theorem). In continuous nonrelativistic cases, $\mathbf{J}_g^{\mathbf{\Omega}}$ have the forms that had earlier been used by Chemists in the so-called Topological Quantum Theory to describe atomic parts ("chemical fragments") of larger units, such as polyatomic molecules – while for Dirac/Weyl or lattice models they appear to have forms that resemble the corresponding relativistic forms (with recent works of this type being on graphene and other Dirac systems that appear in Condensed Matter Physics). And in spite of the fact that the above boundary terms originate from a deep mathematical anomaly (having to do with operators' domains of definitions – an issue that has been briefly studied by a few mathematicians and seems to have been largely ignored by physicists since the beginning of Quantum Mechanics), this note points out examples (from Quantum Condensed Matter Physics) where such non-Hermiticity patterns have physical significance**;** and this seems to cover cases of both conventional and topologically nontrivial materials.

We actually notice examples (with the above non-Hermitian terms acquiring physical significance) mostly in areas such as the so-called Modern Theory of Polarization and of Orbital Magnetization as well as in Applied Physics (where recent work is pointed out on even the off-diagonal version of well-known quantal theorems possessing the associated non-Hermiticities). One may also argue that these non-Hermitian boundary terms can give a concrete example of the bulk-boundary correspondence in topologically nontrivial materials, something however that remains to be seen in detail in future studies. Furthermore, in all stationary state examples considered, these non-Hermitian boundary terms have turned out to be quantized, this quantization being either of conventional (Bohr-type) or of a topological (Quantum Hall Effect (QHE)-type) origin. The latter claim is here substantiated through direct application of Ehrenfest theorem to a simple two-dimensional QHE arrangement (the well-known Landau problem (electron gas in a perpendicular magnetic field) in an external in-plane electric field). Finally, the above non-Hermitian terms are also demonstrated to correct the standard uncertainty relations (of Kennard/Robertson-type) by modifying the uncertainty product in a manner that is consistent with certain well-defined momenta in multiply-connected systems (and in fact they make the correction in a *topologically invariant way* so that the consistency of the uncertainty relations is valid *independent of geometrical details*, as we shall see). Similar results follow for any system that satisfies the Bloch theorem, hence for any spatially periodic system.

The first published report of an example of the above type of paradox in the standard quantum mechanical formalism was ref. [1]. It pointed out (without resolution) an inconsistency in the application of the Ehrenfest theorem (namely the evaluation of the time-derivative ($\frac{d}{dt}$) of the expectation value of an input operator that, in that initiating work, was the position operator in a one-dimensional system). The paradox that was pointed out was an inconsistency (and it is indeed a serious inconsistency, that appears even in contemporary physical applications that involve the standard velocity operator), namely the fact that the expectation value of position (for us now in any dimensionality) $<\Psi(t)|\mathbf{r}|\Psi(t)>$ if evaluated in a stationary state $|\Psi(t)>$ (of a static Hamiltonian) should obviously be independent of time (as the phase factors due to $|\Psi(t)> \sim e^{-iEt/\hbar}$ cancel out), hence we should have

$$\frac{d}{dt}<\Psi(t)|\mathbf{r}|\Psi(t)> = 0, \qquad (1)$$

which however generally contradicts with the standard result that this should be equal to the expectation value of the standard velocity operator $\mathbf{V}$ defined as

$$\mathbf{V} = \frac{i}{\hbar}\,[H, \mathbf{r}], \qquad (2)$$

whose expectation value is generally nonzero (indeed for scattering states – i.e. for a plane-wave state $\Psi(\mathbf{r},t) = Ce^{i\mathbf{k}.\mathbf{r}}\, e^{-iEt/\hbar}$ – it turns out that $<\Psi(t)|\mathbf{V}|\Psi(t)>=|C|^2\hbar\mathbf{k}/m$ that is in general not zero and in fact contains important physical information, namely the global probability flux (quantum mechanical current)), hence a paradox at the very heart of the standard formalism of quantum mechanics.

Historically speaking it is also an important inconsistency, as this later led to further paradoxes associated with the so called Hypervirial theorem in Chemistry (i.e. see the book [2]). Such type of paradoxes (at any dimensionality and with input operators different from $\mathbf{r}$ – also including differential operators as in the well-known Hellmann-Feynman theorem) can always be resolved by retaining some *boundary terms* (after a necessary integration by parts, that is briefly described in what follows for a three-dimensional system, although it can be similarly generalized for any dimensionality). These boundary terms are a reflection of a (hidden) non-Hermiticity of the kinetic energy operator, something that seems not to have been properly emphasized in the literature – providing a partial motivation for this note.

Let us present the main argument and first work out a general three-dimensional example of the application of Ehrenfest theorem with the input operator $\mathbf{\Omega}$ being any vector operator that depends on position ($\mathbf{r}$) and/or canonical momentum ($\mathbf{p}$) operators and that generally has explicit time-dependence. The total time-derivative of the expectation value of $\mathbf{\Omega}(\mathbf{r},\mathbf{p},t)$ is then

$$\frac{d}{dt}\langle\Psi(t)|\boldsymbol{\Omega}|\Psi(t)\rangle = \langle\Psi(t)|\frac{\partial\boldsymbol{\Omega}}{\partial t}|\Psi(t)\rangle + \langle\Psi(t)|\boldsymbol{\Omega}|\frac{d}{dt}\Psi(t)\rangle + \langle\frac{d}{dt}\Psi(t)|\boldsymbol{\Omega}|\Psi(t)\rangle \qquad (3)$$

which by the basic dynamical evolution law $|\frac{d}{dt}\Psi(t)\rangle = \frac{1}{i\hbar} H |\Psi(t)\rangle$ yields

$$\frac{d}{dt}\langle\Psi(t)|\boldsymbol{\Omega}|\Psi(t)\rangle = \langle\Psi(t)|\frac{\partial\boldsymbol{\Omega}}{\partial t}|\Psi(t)\rangle + \frac{1}{i\hbar}\langle\Psi(t)|\,(\boldsymbol{\Omega}\, H - H^{+}\,\boldsymbol{\Omega})\,|\Psi(t)\rangle\,, \qquad (4)$$

and this gives the standard "Heisenberg equation" if one assumes that H is Hermitian ($H^{+} = H$) (and then the last term contains the expectation value of the standard commutator $[\boldsymbol{\Omega}\ ,\ H]$). However, if we allow for possible non-Hermiticities (hence $H^{+} \neq H$) and if we write H=T+U (a kinetic energy and a potential energy operator (assumed real in this note)) and if for simplicity for the moment we ignore the presence of any magnetic vector potentials so that the kinetic energy operator is in position representation just $T= - \hbar^2\, \boldsymbol{\nabla}^2/2m$, we then have (by adding and subtracting $H\,\boldsymbol{\Omega}$) that the above difference can be written as

$$\langle\Psi(t)|\,(\boldsymbol{\Omega}\, H - H^{+}\,\boldsymbol{\Omega})\,|\Psi(t)\rangle = \langle\Psi(t)|\,[\boldsymbol{\Omega}\,,\, H]\,|\Psi(t)\rangle + \langle\Psi(t)|\,(H\,\boldsymbol{\Omega} - H^{+}\,\boldsymbol{\Omega})|\Psi(t)\rangle \qquad (5)$$

the first term reflecting the well-known result (contained in the so-called Heisenberg equation in the standard textbook-literature) and the last term describing the new (and hidden) non-Hermiticity of the Hamiltonian. This last term can then be written as
$-\hbar^2/2m\ (\langle\Psi(t)|\,\boldsymbol{\nabla}^2\,|\boldsymbol{\Phi}(t)\rangle - \langle\boldsymbol{\Phi}(t)|\,\boldsymbol{\nabla}^2\,|\Psi(t)\rangle^{*})$, where we have defined $|\boldsymbol{\Phi}(t)\rangle = \boldsymbol{\Omega}\,|\Psi(t)\rangle$, and this can be evaluated by passing to the position representation**;** it is then equal to the volume integral (over all space) of the quantity $\hbar^2/2m\ (\Psi^{*}\boldsymbol{\nabla}^2\Phi - (\Phi^{*}\boldsymbol{\nabla}^2\Psi)^{*})$ with $\Psi$ and $\Phi=\Omega\Psi$ being the corresponding wavefunctions $\Psi(\mathbf{r},t)$ and $\Phi(\mathbf{r},t)$ (where we have here taken for simplicity an example where $\boldsymbol{\Omega}$ has only one component – for the most general result see application of the above analytical machinery to the physics of orbital magnetization [6]). The resulting expression requires integration by parts in three dimensions, that can be carried out by proper use of the divergence theorem**.** Indeed, from the vector identity

$$\Psi^{*}\boldsymbol{\nabla}^2\Phi - (\Phi^{*}\boldsymbol{\nabla}^2\Psi)^{*} = \boldsymbol{\nabla}\cdot(\Psi^{*}\boldsymbol{\nabla}\Phi - (\Phi^{*}\boldsymbol{\nabla}\Psi)^{*}) \qquad (6)$$

we obtain (with $left\ hand\ side = \Psi^{*}\boldsymbol{\nabla}^2\Phi - (\Phi^{*}\boldsymbol{\nabla}^2\Psi)^{*}$) that

$$\iiint_{all\ space}(left\ hand\ side)\ \ dV = \oiint_S \mathbf{F}\cdot d\mathbf{S} \qquad (7)$$

with the vector field $\mathbf{F}$ defined by $\mathbf{F} = \Psi^{*}\boldsymbol{\nabla}\Phi - (\Phi^{*}\boldsymbol{\nabla}\Psi)^{*}$. We clearly see therefore that the non-Hermitian boundary term can always be written as a flux of some quantity across the system's boundary. If we put all ingredients (i.e. constants) together, then the standard Heisenberg equation is finally augmented by a boundary term, which is the flux (across the system's boundary) of certain generalized current densities $\mathbf{J}_g^{\boldsymbol{\Omega}}$. In the present case with the input operator having a single component $\Omega$, these generalized current densities have the form

$$\mathbf{J}_g^{\Omega} = -\frac{i\hbar}{2m}\,\mathbf{F} = -\frac{i\hbar}{2m}\,(\Psi^{*}\boldsymbol{\nabla}\Phi - (\Phi^{*}\boldsymbol{\nabla}\Psi)^{*}), \qquad (8)$$

always with $\Phi=\Omega\Psi$ (hence they have a form that reduces to the standard quantum mechanical current density $\mathbf{J}= -\frac{i\hbar}{2m}(\Psi^*\nabla\Psi - \Psi^*\nabla\Psi^*)$ whenever $\mathbf{\Omega}$ is the identity operator, namely $\mathbf{J}_g^{\mathbf{1}} = \mathbf{J}$). Such generalized currents have been earlier discussed (with several simple examples in general quantum systems and also in Solid State Physics (most importantly satisfying the Bloch theorem)) in ref. [3] – and they are equivalent to the generalized currents that have been used in Chemistry [4]; they possess the interesting property that they satisfy a continuity equation which however has extra nonvanishing source terms (containing the commutator $[\mathbf{\Omega},H]$), see details in refs. [3],[5]): indeed by also defining a generalized density $\rho_g^{\mathbf{\Omega}} = \Psi^* \,\mathbf{\Omega}\, \Psi$ one has the continuity-type of equation

$$\partial\rho_g^{\mathbf{\Omega}}/\partial t + \nabla\,.\,\mathbf{J}_g^{\mathbf{\Omega}} = \frac{1}{i\hbar}\Psi^* [\mathbf{\Omega},H]\, \Psi \qquad (9)$$

which if integrated over the whole system's volume yields the Ehrenfest theorem (always the "diagonal" version that refers to the expectation value of $\mathbf{\Omega}$) augmented with the non-Hermitian boundary term. Moreover, ref. [5] has gone further than the above "diagonal" cases (namely the standard examination of expectation values in the Ehrenfest theorem) by taking a serious look at the off-diagonal version of this theorem, displaying a number of little surprises (which need to be studied further, in order to reveal their behavioral patterns in a more systematic way, this applying especially to the cases of the off-diagonal Hellmann-Feynman theorem). Furthermore, the recent works [6] and [21] apply the above in the most general setting (for a multi-component $\mathbf{\Omega}$) and in a completely different manner and to more complicated systems of Solid State Physics (in a Bloch theoretic framework), namely in the framework of the so called Modern Theories of Polarization and of Orbital Magnetization (including Thouless's charge pumping); they demonstrate that these non-Hermitian contributions are real (physically relevant), they are not so uncommon, and they carry out important physical information on boundary contributions hidden in certain physical properties such as the polarization and the orbital magnetization of solids (areas – and properties - where the action of the anomalous operator $\mathbf{r}$ is central). It should also be added that these non-Hermitian contributions that can easily be written in closed form (and they can analytically exhibit their behavior in several cases of practical interest (as in refs. [3], [5], [6] and very recently in [21])), are intertwined with Berry's phase physics (Berry curvatures in a parameter space are affected by them – something that needs to be further addressed in the future) and are actually demonstrations of a topological anomaly as pointed out by mathematicians [7]; these anomalies are expected to occur (at least in the diagonal cases above) whenever the input operator $\mathbf{\Omega}$ throws the wavefunctions out of the Hilbert space of the system - see detailed mathematical works in ref. [8] and [9].

Before proceeding it is important to point out that the above paradoxes are directly related to an inconsistency in the standard uncertainty relation in case of systems that move in multiply-connected spaces, i.e. for a quantum particle that moves along a one-dimensional ring (that could be threaded by an Aharonov-Bohm magnetic flux [22]). In such a system, and if the particle is in a definite stationary state with a well-defined canonical momentum, we have a position uncertainty that is finite (of the order of the ring circumference) and a momentum uncertainty that is zero; this gives an uncertainty product that is zero, which seems to violate the (Kennard/Robertson) lower bound $\hbar/2$. This issue is resolved by exactly the non-Hermitian term that is the central concept of the present note (notice, despite the fact that the boundaries do not exist any more – because of a folding and gluing procedure – the non-Hermiticity still remaining): it turns out that the presence of this non-Hermitian term is exactly what is needed in order to give a vanishing uncertainty product – although in the corresponding literature (essentially based on (φ,Lz)-commutation relations) it has not been mentioned as a non-Hermitian correction. For relevant recent work (with the references therein leading to earlier reviews) see ref. [10] and [11] (the latter being one of the rare works that mention the sensitivity of the uncertainty relations to the precise boundary conditions imposed on a system, and also applies the above correction to spatially periodic systems that satisfy the well-known Bloch theorem)**;** the generalized uncertainty product that has also been earlier investigated on a ring in [11], since it refers to p, is like the "square root" of the emergent non-Hermiticity pointed out in the present note (as the procedure followed here involved $< p^2 >$)). Finally, ref. [12] has to be mentioned, where the *topological invariance* of this correction is rigorously shown. The fact that the non-Hermitian term that restores the consistency/correctness of the uncertainty relation (a property at the very heart of Quantum Mechanics) is exactly such that this restoration occurs in a topologically invariant manner may well turn out to be an important general property – and possibly related to the connection with Topology that is hinted in what follows.

Although all the above were concerned with continuous nonrelativistic (Schrödinger) systems (and actually without a magnetic vector potential **A** – hence Aharonov-Bohm type of effects (with the system being outside magnetic fields but enclosing inaccessible magnetic fluxes) as well as cases with nonzero magnetic fields applied on the system – all cases that have seemingly been left out), it is not quite so**:** when there is an extra **A** it is straightforward to generalize the above integrations appropriately, and the analytical form of the generalized currents $\mathbf{J}_g^{\mathbf{\Omega}}$ is adjusted accordingly (the new forms having actually been used in [3], [5], [6] and [21]). But the most interesting generalizations (with few or no applications in the literature so far) occur **(a)** in the case of continuous Dirac/Weyl systems (see i.e. [19], [20] for application to graphene and other Dirac materials), in which cases $\mathbf{J}_g^{\mathbf{\Omega}}$ contain the Pauli operators $\boldsymbol{\sigma}$ in place of the del operator $\boldsymbol{\nabla}$ in their definition, and **(b)** in the case of discretized systems, such as lattice models (that routinely appear in Solid State Physics through a tight-binding approximation) where the above theory of non-Hermitian boundary terms needs to be discretized, i.e. in the spirit of refs [13], [14] (works that are applicable to discretizations of non-Hermitian models). Such generalizations (of the forms of $\mathbf{J}_g^{\mathbf{\Omega}}$) are urgently needed, as such continuous or discrete (pseudo-) relativistic models nowadays appear in a large number of works (and their number and importance keeps increasing, but almost always referring to Hermitian kinetic energy models) because of the recent explosion due to graphene, topological insulators and superconductors, and Dirac and Weyl semimetals (for an overview of the standard Hermitian physics of these systems see refs [15] and [16], and for the newly discovered non-Hermitian generalizations see [25] and [26]).

Let us also reiterate a couple of remarks that deserve further attention**:** there is evidence of *quantization* of all these non-Hermitian fluxes (at least for stationary states), that can be of a conventional (Bohr-type) origin (as in an Aharonov-Bohm ring [3]) or can be of a topological (Quantum Hall Effect (QHE)-type) origin (i.e. the quantization of boundary forces in ref. [17]). The former type originates from the very resolution of the original paradox emphasized in the beginning of the present note, namely the fact that the non-Hermitian term cancels out the expectation value of the standard velocity operator **V** (eq.(2)) so that it gives the expected zero of eq.(1), and as the <**V**> is quantized (a la Bohr, so that an integer number of half de Broglie wavelengths fits into the circumference) so is the non-Hermitian term as well. Similarly, if we consider a 2D plane in an external perpendicular magnetic field (we call it the Landau problem) and in the additional presence of an in-plane electric field, a problem that is completely solvable (in some Landau gauge), it yields in closed form all eigenfunctions, energies (the well-known *tilted* Landau Levels) and the global probability current (which is nonzero due to the tilting, hence due to the removal of the usual Landau Level degeneracy). If then to this system we apply the Ehrenfest theorem in 2-D, for the coordinate that is parallel to the edges (or parallel to the direction where we can apply periodic boundary conditions, which is also normal to the direction of the electric field), then a similar cancellation-argument as in the beginning of this note leads to quantization of the non-Hermitian term. The latter type (similar to the one that has been noticed in ref. [17]) that seems to be of a topological nature, can be seen through the Ehrenfest theorem again, but now with the *momentum* as the input operator. In such cases it has been argued (refs. [6] and [21]) that the direct connection between these non-Hermitian boundary terms with the corresponding bulk quantities is actually a reflection of the well-known bulk-boundary correspondence [18] in topologically nontrivial materials. However, such mathematically esoteric issues (together with the possible use of these non-Hermiticities as more or less practical tools in describing the well-known dissipationless boundary (edge-) states in topological materials) are still open and require a more focused study.

With respect to the above type of Landau problems, it should also be pointed out that application of our emergent non-Hermitian boundary term in the case of guiding center coordinates $X_0,Y_0$ (in the Quantum Hall Effect system) would give a *generalized* uncertainty relation for the uncertainty product $\Delta X_0\ \Delta Y_0$ (a correction of the right-hand-side of the standard $\Delta X_0\ \Delta Y_0 \geq l_B^2/2$ with $l_B$ the well-known magnetic length). This would finally give zero for the right-hand-side, thereby ageeing with the Landau solution of the Landau problem (in a Landau gauge) where we have a sharply defined i.e. $y_0$ and periodic boundary conditions in the x-direction. In this case, due to the folding in the x-direction, we have a cylinder (the one that Laughlin used for his well-known gauge argument that explains the Integer Quantum Hall Effect), in this case a vertically placed one, and because of the finite length in the x-direction, the above uncertainty $\Delta X_0$ is finite (not infinite), whereas $\Delta Y_0=0$ (as $y_0$ is precisely-defined). So, the above uncertainty product is indeed zero – and this is explained by the generalized uncertainty relation (that includes the non-Hermitian correction). [This could probably be viewed as the quantum analog of the semiclassical skipping orbits usually invoked in standard discussions in this area.] Note that this sharpness of $y_0$ is absolutely necessary, in order for the sharpness (or precise quantization) of the Hall conductance (as shown by the Laughlin argument). Hence it seems that the emergent non-Hermitian term (although its initial origin, the boundary, has disappeared due to the folding) is an absolute must to be considered, in order for the IQHE to be explained. [And it is plausible to wonder whether this need (of consideration of the emergent non-Hermiticity) might be generalizable to all topological effects.]

Reference should also be given to recent work [27] where a unified recipe showing a conserved current (of a non-conserved quantity) is discussed whose circulation characterizes the corresponding orbital magnetization and whose net flow vanishes at equilibrium. (One can see our non-Hermitian term – the vanishing of the net flow actually having to do with our above mentioned *cancellations* – and then, this recent preprint gives various different results that can actually be viewed as physical interpretations for our non-Hermitian terms (mainly for the spin-torque (preprint's ref. 6), and concepts related to superconductivity (i.e. charge current renormalized by the condensate backflow due to the coupling between quasiparticles and condensate (preprint's ref. 22)). Also a Berry phase formula for the orbital magnetization in superconductors is found, which consists of both the local and global charge of quasiparticles. In other areas discussed in this preprint there is the issue of the conserved current of non-conserved quantities, such as: magnon and phonon systems described by bosonic Bogoliubov-deGennes Hamiltonians that is of importance for studying the spin Nernst effect of magnons in a noncollinear antiferromagnetic insulators [preprint's refs. 50, 51] and also due to magnon-phonon interactions in collinear ferrimagnets [preprint's ref. 52] as well as the phonon angular momentum Hall effect [preprint's ref. 53] – and in superconductors, this steady-state semiclassical theory needs to be extended to include time derivatives and gauge fields in order to describe gauge-invariant coupled dynamics of quasiparticles and condensate in the presence of electromagnetic fields. All this can quite possibly be seen under our magnifying glass, with emergent non-Hermiticities lurking around – although this also needs a more focused study.

It is probably more important to emphatically point out that what we have presented is actually compatible with the so called skin effect in general non-Hermitian systems in Solid State Physics (with tremendous activity in the last couple of years in solid state models - i.e. for the most recent work see [23]), and it is expected to be directly related to exact solutions of non-Hermitian (tight-binding) models *with non-reciprocal hopping* under generalized boundary conditions. The present note focuses on the original *continuous* models (without discretization approximations, routinely carried out in Solid State tight-binding models). It would then be marvelous to connect (or map our continuous formulations of) these emergent non-Hermiticities to the many available discretized models (where many new results, that go beyond the standard bulk-boundary correspondence - in topological systems - are currently available, some of them leading to the so-called exceptional points where eigenstates and eigenvalues coalesce). One would expect that the connection (mapping) to our continuous system would lead to an enhanced understanding of a multitude of issues of fundamental importance (at least in non-Hermitian Condensed Matter Physics).

Finally, it should be reiterated that, usually, we think of non-Hermitian behaviors as appearing only in open systems (they are actually deviations from the axiomatically imposed mathematical theory of Quantum Mechanics of closed systems), and that their interaction/exchange of information with the environment is the one that is responsible for the non-Hermiticity [here references should be given to the many non-Hermitian works, starting with the seminal 1998 paper of Bender and colleagues [28] and also of a couple of books [29] and reviews [25,26] on "non-Hermitian Physics"]. In contrast to all these works that are already forming an entirely new area of physics, here we see that, if a system is multiply-connected (for example it is folded along one Cartesian direction, such as a ring, a cylinder, a torus etc.), there is the peculiarity that, in spite of being a closed system, it can be treated *like* an open one (with the "environment" being identical to the system itself). In addition, note that the generalized current (based on the position operator) that we have stressed above (as carrying the information of the non-Hermiticity) *senses* the position (by its definition, eq. (8), with the input operator $\mathbf{\Omega} = \mathbf{r}$), and because of the folding, the point of glue, although physically being a single point, is in two different points along this direction, hence it has 2 different values (a discontinuity), thus leading to the nonzero flux of this position-generalized current, which finally gives the non-vanishing (and non-trivial) non-Hermiticity. Here, important concepts might be hiding (that may have a general relativistic

character – as this position-generalized current can be viewed as a "flow of position"), and this non-Hermiticity that has remained even after the removal of the boundaries (due to the gluing procedure) might prove to be a very fundamental common thread in all of Physics.

---